\documentclass[runningheads]{llncs}
\usepackage{graphicx}
\usepackage{comment}
\usepackage{todonotes}
\usepackage{physics}
\usepackage{makecell}
\usepackage{enumitem}

\begin{document}
\title{Parsec: A State Channel for the Internet of Value}
%
%
\author{Amit Kumar Jaiswal\orcidID{0000-0001-8848-7041}}
\authorrunning{Jaiswal}
%
\institute{University of Bedfordshire, Luton, UK \\
\email{amitkumar.jaiswal@beds.ac.uk},\\ }

\maketitle              
\begin{abstract}
We propose Parsec, a web-scale State channel for the Internet of Value to exterminate the consensus bottleneck in Blockchain by leveraging a network of state channels which enable to robustly transfer value off-chain. It acts as an infrastructure layer developed on top of Ethereum Blockchain, as a network protocol which allows coherent routing and interlocking channel transfers for trade-off between parties. A web-scale solution for state channels is implemented to enable a layer of value transfer to the internet. Existing network protocol on State Channels include Raiden for Ethereum and Lightning Network for Bitcoin.
However, we intend to leverage existing web-scale technologies used by large Internet companies such as Uber, LinkedIn or Netflix. We use Apache Kafka to scale the global payment operation to trillions of operations per day enabling near-instant, low-fee, scalable, and privacy-sustainable payments. 
Our architecture follows Event Sourcing pattern which solves current issues of payment solutions such as scaling, transfer, interoperability, low-fees, micropayments and to name a few. To the best of knowledge, our proposed model achieve better performance than state-of-the-art lightning network on the Ethereum based (fork) cryptocoins.
\end{abstract}
%
%
\section{Introduction}

The influence of data systems in finance has been so far focused on cutting administrative costs. However, technology could transform business models in finance. With the help of cryptography on distributed data systems, we can design infrastructure that not only keeps track of transactions and balances but also allows transfer of assets and enforcement of contracts without the need of clearinghouses, middle men, escrows and even some tasks performed by lawyers. On these systems settlements are final and there is no possibility of repudiation while smart contracts execute deterministically according to predefined inputs. Institutions that enclasp this vision will have a competitive advantage. Investing in distributed cryptographic technology and regulatory oriented technology will pay off in the form of less intermediation, less legal fees or exordium and ultimately an accelerated innovation cycle where new financial services can be provided with much less friction and where these services are much better tailored to smaller market segments. These institutions will be able to compete in the long tail of financial services and offer a much more competitive service for the mainstream.

However, for the Internet of Value vision to happen, there’s need for a whole redesign of the architectural Information Technology (IT) principles used to build new services. In this paper, our proposed models can achieve better performances than state-of-the-art existing models on the Ethereum Blockchain~\cite{eth_block} network.

\section{Architectural Details}
\begin{figure}[h!]
	\centering
	\includegraphics[width=0.8\linewidth]{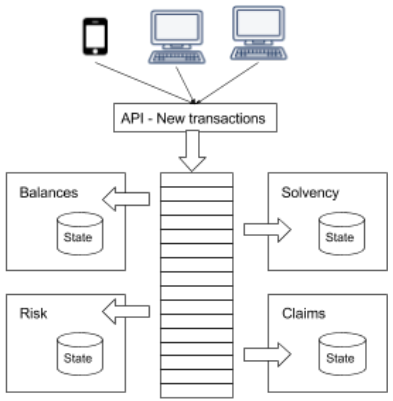}
	\caption{Architecture to create new services}
	\label{fig:w1}
\end{figure} 
In search of new services such as global payment solution, we created an architecture that makes possible to quickly innovate and create new services. In order to achieve this at web scale, we attempt to employ the Event Sourcing~\cite{event_sourcing} pattern, an approach to handle large operations on data driven by a series of events, in which each event is recorded in an append-only datastore. In this kind of architecture, instead of having a centralized state for all the applications in a centralized database or a set of disconnected state holding databases, we rely on a centralized log of state changes, which allows to create new applications very quickly because it is only necessary to replay the log of changes in order to build new services with new states that will be consistent with the state being held by other services in the same company.

\textit{Event Sourcing} is very powerful but has a coupling
problem. In big organizations a central team needs to own the event log.
\begin{figure}[h!]
	\centering
	\includegraphics[width=0.8\linewidth]{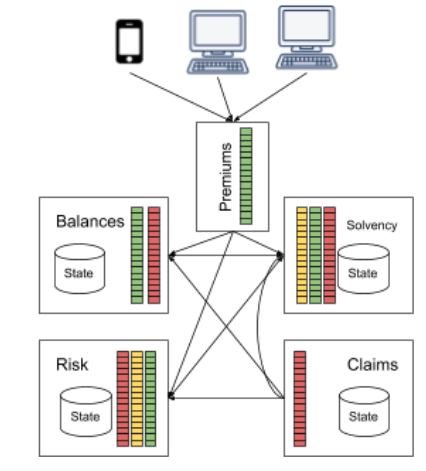}
	\caption{Architecture based on Event Sourcing Pattern}
	\label{fig:w2}
\end{figure} 
Usually, many different teams want to access the log and they all need to agree on the technology and permissions to write and read from it. Also teams reading or writing intensively from or to the central infrastructure risk causes performance problem to the whole.

Also, this kind of architecture is not suitable for collaborating among different organizations or group of companies. For that reason, an architecture where each business unit or application would hold a copy of the parts the log relevant to it would be better. If the stream of data is composed of different topics (for example, premium payments or claims), then each business unit would be able to hold a copy of the topics relevant to it. The above
architecture would be feasible in a trusted environment but in a collaborative environment, all transactions would have to be signed
by the different parties and the order of the transactions should be guaranteed.

We can achieve that in a manner similar to blockchain by means of a linked stream of events using hash pointers to the previous events and by embedding signatures of both sides of a transaction in the message itself.
\begin{figure}[h!]
	\centering
	\includegraphics[width=0.8\linewidth]{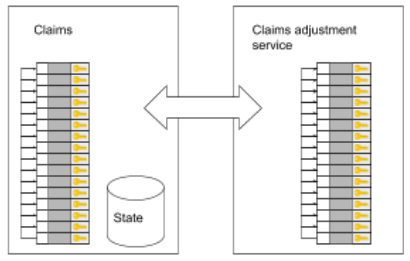}
	\caption{Sub-model based on Linked stream of events}
	\label{fig:w3}
\end{figure} 
This would imply that some business units would have to act as signing authorities to their counter-parties in external organizations. By acting like that we would achieve the benefits of a distributed ledger but without being able to transfer assets.

Asserting that assets can’t be transferred with only a distributed ledger can be contentious. Indeed if both parties have proof that the transaction has been agreed.

However, even in this case where the transaction implied a transfer of assets, for example, in the case where the transaction implied a transfer of assets, for example, in the case of signing an insurance policy, this would not imply that any of the parties would pay the stipulated amounts. We would have no guarantee that policy takers would pay the claims. For that purpose we would need to resort to a third party escrow or even better to a Smart Contract running in a
blockchain with distributed consensus. It is arguable whether this blockchain would have to be public or if a permissioned distributed ledger would be able to achieve this purpose.
\begin{figure}[h!]
	\centering
	\includegraphics[width=0.8\linewidth]{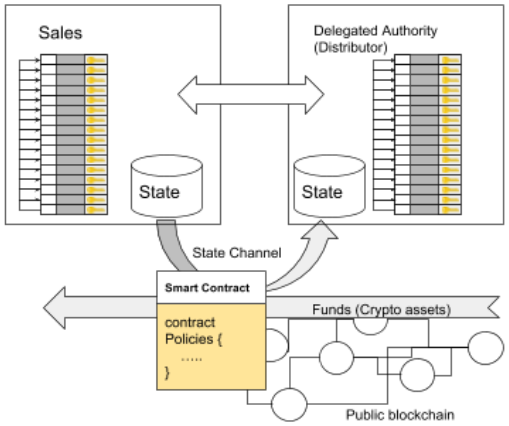}
	\caption{Model for State Channel and Smart
Contract operation on Public Blockchain}
	\label{fig:PCA_WIKI}
\end{figure} 

Here we assume it as a public blockchain capable of running Smart Contracts. In fact the structure we described here is following an existing approach, so called a State Channel~\cite{statechannel} and it allows untrusted parties to exchange transactions with the possibility to resort to an independent escrow held in a Smart Contract operating on a public blockchain in case that one of the parties does not fulfill its commitment.
\begin{figure}[h!]
	\centering
	\includegraphics[width=0.8\linewidth]{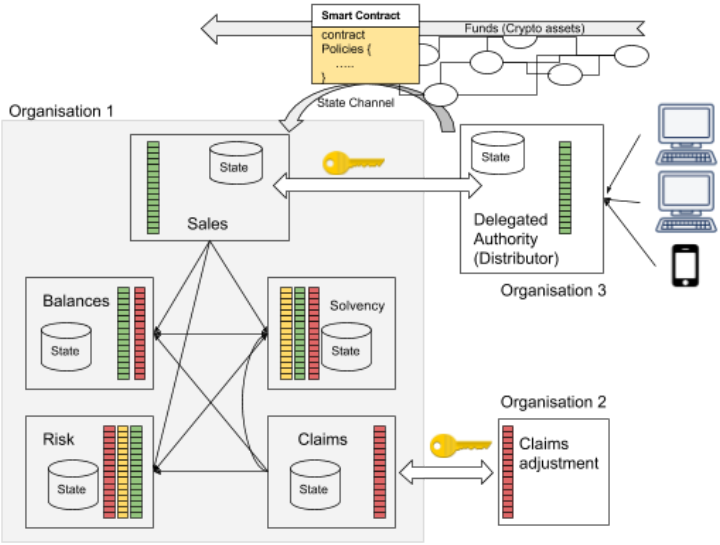}
	\caption{Architecture of Parsec Infrastructure Layer}
	\label{fig:w5}
\end{figure}

Finally get the big picture on how a state of the art architecture for a financial system should look like. It would have to: (1) Be based on transaction logs, (2) It would have to include a replication mechanism of such logs between business units, (3) Logs would have to be enhanced to become shared ledgers when dealing with external untrusted parties, and (4) State channel support would have to be added for asset transfer.

\section{Implementation}
We implemented a infrastructure layer named Parsec\footnote{Code Available at: https://github.com/quanonblocks/Parsec} using Apache Kafka, a distributed data system that can be used as an event sourcing unified log tool. It can be seen both as a queue or as a data store. We intend to implement only the very minimum additions to the already existing functionality based on~\cite{uber,LinkedIn,netflix} so that it can work as a State Channel. For that purpose a fundamental tool of the Apache Kafka~\cite{kafka} ecosystem will be Kafka Streams~\cite{kafka_stream} as it already offers many of the functionalities we need to implement.

We integrated few major functions on top of state channel to work which are (1) Scalability and replicability is provided by Apache Kafka and Kafka mirror maker, (2) Ability to maintain state at scale and interaction with Smart Contract or chaincode~\cite{chaincode} is extended by Parsec on top of Kafka Streams, and (3) Guarantees an exact sequence of hash pointers and transaction verification with its cryptographic signatures is developed in Parsec.

For above given functionality, we intend to provide a library running on top of Kafka Streams under fast data solution system Landoop.
Streams that guarantees the last three points.
\subsection{Parsec Nodes}
Parsec nodes are the job instances of the Parsec Library implementing the Parsec Software Development Kit (SDK). The Parsec library is developed on top of the Kafka streams library with the
several addition of functionality such as support for a catalogue of various kinds of transactions and to keep an update state for all unique Ids of the transaction topic (kafka topic). Also, we have integrated the Hashed Timelock~\cite{hashed_time} to interact with Ethereum smart contracts~\cite{smart_contract} for the kafka topic partitions under the nodes' supervision for scheduled settlements or the request of control topic.

All functionality for each Parsec job relates only to the partitions under it’s control.

\subsection{Parsec Transactions}
A basic catalogue of predefined transactions is implemented which can be used to add additional custom transactions.

Micropayment transaction~\cite{micropayment} performs the following actions:
\begin{itemize}
  \item Keep a sliding array of the latest n number of transactions in order to reconstruct the exact order of the transactions from the transaction hash pointers. This is performed in order to cope with our order of messages or repeated messages.
  \item Calculate the balance after each transaction.
  \item Commit the balance to the Smart Contract in the Ethereum network at predefined intervals. Intervals are defined by the modulo m of the incremental transaction sequence number.
  \item Commit the balance to the smart contract in the Ethereum network at predefined timeouts. The timeout duration can be specified by the \textit{checkpoint\_timeout} parameter.
  \item Under the request of the control topic, perform an unscheduled settlement request to the smart contract.
  \item Under the request of the control topic, perform a disputed settlement request to the smart contract by means of transferring the signed transactions since the last settlement. 
\end{itemize}

\section{Conclusion \& Future Works}
This paper attempts to address the challenge in scaling, fast transaction processing, verification process issues simplifies with our proposed model contains Parsec Simulator which generates a special protocol messages under Kafka cluster with fully functional API support for existing cryptocurrencies like Bitcoin~\cite{bitcoin} and Ethereum~\cite{ethereum} accessible via Avro User Interface{\footnote{Avro UI: https://github.com/kaaproject/avro-ui}}. We also have supports to check Schema's~\cite{schemas} and protocol messages with Coyote test suite{\footnote{Coyote Tester: https://github.com/Landoop/coyote}}. For future work, we are looking forward to provide additional ways to support State Management functionality.

\section*{ACKNOWLEDGMENT}
This work is supported by Quanonblocks LLP.

\section*{Appendix}
\subsection*{Parsec Protocol}
The Parsec protocol introduces a layer for cryptocurrency transactions to take place, without necessarily being registered into the global blockchain.
\begin{figure}[h!]
	\centering
	\includegraphics[width=0.8\linewidth]{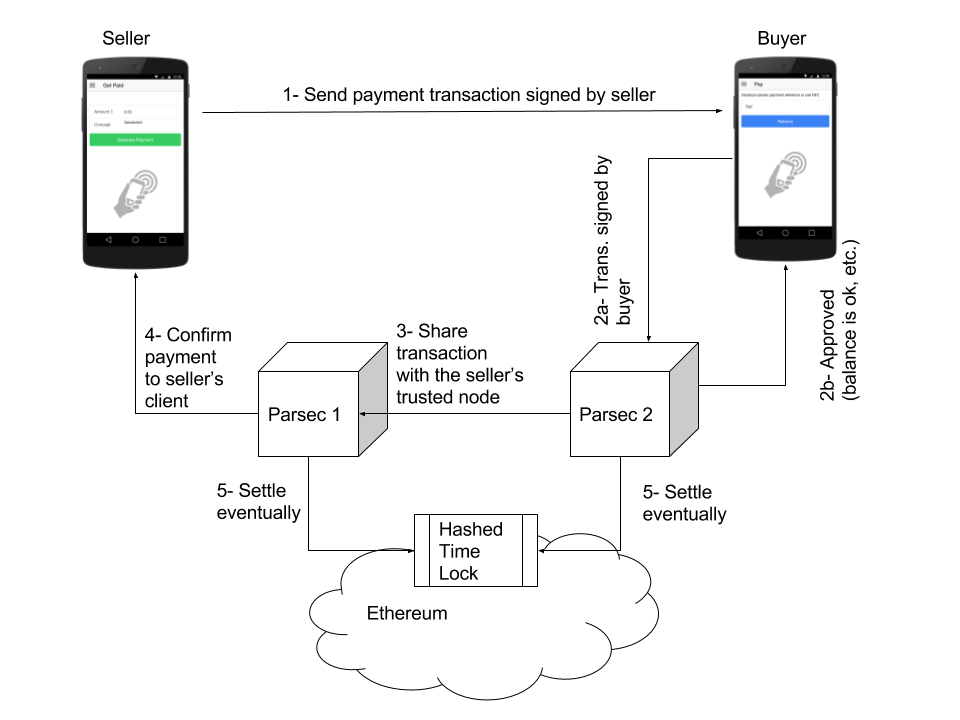}
	\caption{Customer Journey via Parsec Protocol}
	\label{fig:customer}
\end{figure}

\medskip

\noindent
{\it Parsec Protocol}
\begin{verbatim}
package org.parsec.protocol

  val customerPaymentJourney = .stripMargin
  val DEFAULT_CHANNEL = "default"
  val ALLOWED_CURRENCIES = Set("ETH", "BTC")

  /**
    * The seller needs to produce an INVOICE (through Parsec channel)
    *
    * @param invoiceAddress the sellect address for the currency to be transferred to
    * @param price          the amount of currency to exchange
    * @param currency       ETH or BTC supported
    * @param invoiceType    (optional) use to pass-in other metadata about the invoice
    */
  case class Invoice(invoiceAddress: String, price: Double, 
  currency: String, invoiceType: String = "") {
    assert(ALLOWED_CURRENCIES.contains(currency), s"ParsecProtocol: Currency 
    $currency is not an allowed currency. Use: " + ALLOWED_CURRENCIES.mkString(" or "))

    def produceInvoiceHash = "######-####-####"
  }

  /**
    * Here the Parsec Protocol lies here!
    *
    * We need to ensure that the order of Signed Invoices is guaranteed
    * No cluster of distributed systems can guarantee that 
    (due to varying network latencies etc.)
    *
    */
  case class HashPointer(transactionID: String, 
  // i.e. invoiceID
                         transactionHash: String)

  case class SignedInvoice(invoiceID: String = "auto-generate-random-UUID+channel", 
  // import java.util.UUID
                           invoiceSignature: String,
                           supplierAddress: String, // no need to know Private Key
                           buyerPublicKey: String,
                           buyerAddress: String,
                           currency: String,
                           price: Double,
                           channel: String = DEFAULT_CHANNEL,
                           hashPointerToPrevious: HashPointer) {
    assert(ALLOWED_CURRENCIES.contains(currency), s"ParsecProtocol: 
    Currency $currency is not an allowed currency. Use: " + 
    ALLOWED_CURRENCIES.mkString(" or "))
    assert(isValidBuyerAddress)

    def isValidBuyerAddress: Boolean = {
      if (currency == "ETH")
        return buyerAddress equals ethereumAddressFromPublicKey(buyerPublicKey)
      else
        return buyerAddress equals bitcoinAddressFromPublicKey(buyerPublicKey)
    }
  }

}
\end{verbatim}
\noindent
\end{document}